\documentclass[10pt,letterpaper]{article}
\usepackage[top=0.85in,left=2.75in,footskip=0.75in]{geometry}

\usepackage{amsmath,amssymb}

\usepackage{changepage}

\usepackage[utf8x]{inputenc}

\usepackage{textcomp,marvosym}

\usepackage{cite}

\usepackage{nameref,hyperref}

\usepackage[right]{lineno} 

\usepackage{microtype}
\DisableLigatures[f]{encoding = *, family = * }

\usepackage[table]{xcolor}

\usepackage{array}

\newcolumntype{+}{!{\vrule width 2pt}}

\newlength\savedwidth

\raggedright
\usepackage[aboveskip=1pt,labelfont=bf,labelsep=period,justification=raggedright,singlelinecheck=off]{caption}

\makeatletter
\renewcommand{\@biblabel}[1]{\quad#1.}
\makeatother

\usepackage{lastpage,fancyhdr,graphicx}
\usepackage{epstopdf}
\fancyheadoffset[L]{2.25in}
\fancyfootoffset[L]{2.25in}
\usepackage{comment}

\begin{document}
\vspace*{0.2in}

\begin{flushleft}
{\Large
\textbf\newline{Evolution of global development cooperation: An analysis of aid flows with hierarchical stochastic block models}
}
\newline
\\
Koji Oishi\textsuperscript{1,2*},
Hiroto Ito\textsuperscript{3},
Yohsuke Murase\textsuperscript{4},
Hiroki Takikawa\textsuperscript{5},
Takuto Sakamoto\textsuperscript{3}
\\
\bigskip
\textbf{1} Department of International Politics, Aoyama Gakuin University, Tokyo, Japan
\\
\textbf{2} Japan Society for the Promotion of Science, Tokyo, Japan
\\
\textbf{3} Graduate School of Arts and Sciences, The University of Tokyo, Tokyo, Japan
\\
\textbf{4} RIKEN Center for Computational Science, Kobe, Japan
\\
\textbf{5} Graduate School of Humanities and Sociology, The University of Tokyo, Tokyo, Japan
\\
\bigskip

%
%





* oishi@sipec.aoyama.ac.jp

\end{flushleft}
\section*{Abstract}
Despite considerable scholarly attention on the institutional and normative aspects of development cooperation, its longitudinal dynamics unfolding at the global level have rarely been investigated. Focusing on aid, we examine the evolving global structure of development cooperation induced by aid flows in its entirety. Representing annual aid flows between donors and recipients from 1970 to 2013 as a series of networks, we apply hierarchical stochastic block models to extensive aid-flow data that cover not only the aid behavior of the major OECD donors but also that of other emerging donors, including China. Despite a considerable degree of external expansion and internal diversification of aid relations over the years, the analysis has uncovered a temporally persistent structure of aid networks. The latter comprises, on the one hand, a limited number of major donors with far-reaching resources and, on the other hand, a large number of mostly poor but globally well-connected recipients. The results cast doubt on the efficacy of recurrent efforts for ``aid reform'' in substantially changing the global aid flow pattern.

\section*{Introduction}
Over the past 60 years, international society has sustained an increasingly institutionalized pattern of development cooperation. A countless number of programs, funds, and institutions have been launched for promoting economic and social development in different countries and regions; various landmark policy documents (e.g., Berg Report, Comprehensive Development Framework, Paris Declaration) have been adopted for more efficient and effective development efforts; and an ever-expanding array of goals and targets, including the Millennium Development Goals and Sustainable Development Goals, have been globally embraced for channeling these efforts to all countries and all people in need. Not surprisingly, these conspicuous aspects of development cooperation have attracted the attention of international relations (IR) researchers, especially those working in the liberal (e.g., international regime) and constructivist (e.g., aid norm diffusion) traditions \cite{brown2020rise,fukuda2011theory,heldt2019explaining,hook2016development,lumsdaine1993moral}.

However, the actual dynamics of cooperation unfolding at the global level have rarely been investigated. These dynamics have emerged and evolved over years under successive policy and normative initiatives, such as those mentioned above. Our study conducts such an investigation to fill this research gap. Focusing on one of the most traditional policy instruments for development cooperation, that is, aid, the study aims to elucidate the evolving global structure of development cooperation induced by aid flows among various international actors (e.g., states, international organizations, and major public and private funds) over the past decades. For this purpose, we employ cutting-edge methods, including stochastic block models (SBMs), which have been developed in complex network science \cite{Barabasi2016}, and we apply these analytical tools to extensive aid-flow data that cover not only the aid behavior of Organisation for Economic Co-operation and Development (OECD) countries but also that of other emerging donors, including China.

Network science itself has made significant inroads into IR in recent years\cite{hafner2006power,hafner2009network}. Different aspects of IR have been represented as networks of co-affiliation to intergovernmental-organization \cite{beckfield2010social,cranmer2015kantian,greenhill2017clubs}, democratic peace \cite{cranmer2015kantian}, trade \cite{cranmer2015kantian}, voting behavior in the United Nations \cite{pomeroy2019multiplex}, and bilateral cooperation agreements \cite{pomeroy2019multiplex}, among others \cite{de2018using,kim2020global}. Unfortunately, the structure of global development cooperation, including aid, has received only scant attention in this context. There are indeed a few notable studies that employ network analysis to study aid \cite{swiss2016world,swiss2016membership,swiss2017foreign,peterson2011foreign,gutting2017donor}. Many of these studies calculate node-level local measures (e.g., degree centrality and eigenvector centrality) from given network representations, and then input these quantities into regression equations that largely model state-level behavior and/or state-to-state dyadic relationships. Although such a methodological orientation is fully consistent with the dominant literature on aid relations and aid effectiveness in IR \cite{bermeo2017aid,dietrich2016donor,gehring2017aid,krasner2014improving,wright2010politics}, the aid networks represented therein contain far more information than these local quantities can capture. By fully leveraging this information, it is possible to rigorously measure and analyze the entire structure of aid relations unfolding at the systemic level. By performing such measurement and analysis, we aim to demonstrate the still-unexplored potential of network science for studying global development cooperation.

\section*{Materials and methods}
\subsection*{Foreign aid networks}
We begin by building foreign aid networks using open datasets provided by the OECD \cite{OECD2018} and AidData \cite{tierney2011more,dreher2021aid}.
Both datasets track foreign aid provided by both bilateral and multilateral donors.

The OECD dataset on foreign aid is based on reports provided by donor countries, which include the names of donors and recipients, years, and the amount of aid.
We focus on official development assistance (ODA) commitments and do not include other financial flows in our analysis.
ODA needs to satisfy several criteria for public development aid, which are set by the Development Assistance Committee (DAC) of the OECD \cite{OECD2018}.

The OECD data do not cover several emerging donors (e.g., China), but AidData collects a similar set of information (donors and recipients, years, and amount of aid) for both emerging and traditional donor countries \cite{tierney2011more}.
Therefore, we also use AidData for the donors that are not included in the OECD data but appear in AidData.
Data from AidData are based on not only the official reports of donors to OECD but also their original data collection.
In particular, the information about foreign aid provided by China is separately collected based on various information sources, including media reports \cite{dreher2021aid}.
Although China hardly follows the DAC criteria, we include the financial flow that have a similar nature to ODA, which are labeled as ``ODA-like'' by AidData.
Chinese ``ODA-like" assistance as a donor is included only after 2000. In addition, while data for DAC donors are reported and recorded from an earlier period, it is likely that the quality of data for non-DAC donors other than China has also improved. This suggests that data quality improves over time, which may affect the evolution of the aid networks. Therefore, while we must accept the possibility that our analysis may be biased due to data limitations, OECD.stat and AidData are still the sources that many studies rely on, and the results of our analysis maintain some validity.

We construct a weighted directed network by aggregating the amount of aid in the dataset for each year in 1970--2013.
The networks, which we call foreign aid networks (FANs), represent the yearly financial flow between the actors, including  governments, intergovernmental organizations (e.g., the World Bank), and private foundations (e.g., the Bill \& Melinda Gates Foundation).
Specifically, we define the weight of an edge from node $i$ to $j$ as the normalized quantity $f_{ij}/s^{\text{out}}_{i}s^{\text{in}}_{j}$, where $f_{ij}$ is the amount of aid $i$ gave to $j$, $s^{\text{out}}_{i}$ is the total amount of aid $i$ gave, and $s^{\text{in}}_{j}$ is the total amount of aid $j$ received in the year.
This normalization is introduced because an analysis of the network without the normalization yields results that reflect only the financial flows between a few major players and largely ignore many others, as the amount of aid flow $f_{ij}$ is distributed over a wide range of magnitude.
The weight of an edge from a donor to a recipient is equal to one, the maximum, when the donor provided aid only to the recipient and the recipient obtained aid only from the donor in the year.
Even if the amount of aid from a donor to a recipient was large, the edge weight would be small when the donor provided (the recipient obtained) a large amount of aid to other recipients (from other donors) in the year.

\subsection*{Stochastic block models}

In addition to the local network attributes, such as average degree,
we identify larger-scale structures in networks by dividing nodes into communities or blocks.
While there are diverse ways to identify communities in networks~\cite{fortunato2010community}, one of the most principled approaches to performing this task is inferring the parameters for SBMs; this is arguably one of the simplest generative models incorporating the idea of groups of nodes \cite{peixoto2019bayesian}.
The SBM is a simple generative model for random networks, where nodes are partitioned into $B$ blocks $\boldsymbol{b} = \{b_i\}$
and the edges between the blocks are generated according to the parameter $\boldsymbol{\theta}$, where $\theta_{rs}$ is the probability of forming edges between block $r$ and $s$.
Given these constraints, the edges are then placed randomly.
Hence, nodes that belong to the same block possess the same probability of being connected with other nodes of the network.

It is possible to calculate the probability of generating a network $\boldsymbol{A}$ by the SBM with $\boldsymbol{b}$ and $\boldsymbol{\theta}$, $P(\boldsymbol{A}|\boldsymbol{\theta}, \boldsymbol{b})$, in an analytic form.
Therefore, if we observe an empirical network $\boldsymbol{A}$, the posterior probability that it was generated by a given partition $\boldsymbol{b}$ is obtained by
\begin{equation}\label{eq:posterior}
    P(\boldsymbol{b}|\boldsymbol{A}) = \frac{\sum_{\boldsymbol{\theta}} P(\boldsymbol{A} | \boldsymbol{\theta}, \boldsymbol{b} ) P( \boldsymbol{\theta}, \boldsymbol{b} ) }{ P(\boldsymbol{A}) },
\end{equation}
where $P(\boldsymbol{\theta}, \boldsymbol{b})$ is the prior probability of the model parameters and $P(\boldsymbol{A}) \equiv \sum_{\boldsymbol{\theta},\boldsymbol{b}} P(\boldsymbol{A}| \boldsymbol{\theta},\boldsymbol{b}) P(\boldsymbol{\theta}, \boldsymbol{b})$ is the total probability of observing the empirical network $\boldsymbol{A}$.
By finding a network partition that maximizes Eq.~(\ref{eq:posterior}), we can infer the most likely partitions of empirical networks or sample the partitions from the posterior distribution.
Note that the block structure does not always correspond to communities.
On the one hand, communities are sets of nodes that are more densely connected inside of communities compared with outside.
On the other hand, the SBM covers a broader class of structures, such as core-periphery structures, in which nodes in core blocks are likely to connect with nodes in both core and periphery blocks but nodes in periphery blocks tend to connect only to nodes in core blocks.

As FANs are weighted networks, we use the extended version of the SBM, which treats edge weights as covariates sampled from some distribution conditioned on the node partition~\cite{aicher2015learning,peixoto2018nonparametric}.
In other words, the weight of all the edges follows the same probability distribution if the edges connect the same directed pair of blocks (e.g., from block $r$ to block $s$)
We try a couple of probability distributions, including exponential and normal distributions, and find that the log-normal distribution best fits our dataset.
This is because blocks in FANs inferred with normal and exponential distribution tend to include only a single or a few nodes; thus we could not find any meaningful division of FANs.
Therefore, the results for the log-normal model are shown below.

In addition, we use a hierarchical version of the model, which replaces the prior by a nested sequence of priors and hyperpriors, as described in~\cite{peixoto2018nonparametric}.
In other words, the nodes are partitioned into blocks, and each of these blocks is partitioned into higher-order blocks, and so on.
Without allowing the hierarchical structure, the SBM could fail to find small blocks compared with the size of networks.

Partition and other parameters, including the number of blocks and layers in a hierarchy, are determined by optimizing minimum description length (MDL), which measures the amount of information required to describe the data.
A smaller MDL means that we need less additional information to describe the network and indicates that the assumption of block structure is suitable to explain the network.
The inferences of the block structures are conducted by graph-tool \cite{graph-tool}, which is one of the standard libraries for network analysis.
The source code for the analysis is available online to enable reproducibility of the results.

\section*{Results}

\subsection*{Local network characteristics}
\begin{figure}[!h]
  
  \includegraphics[width=14cm]{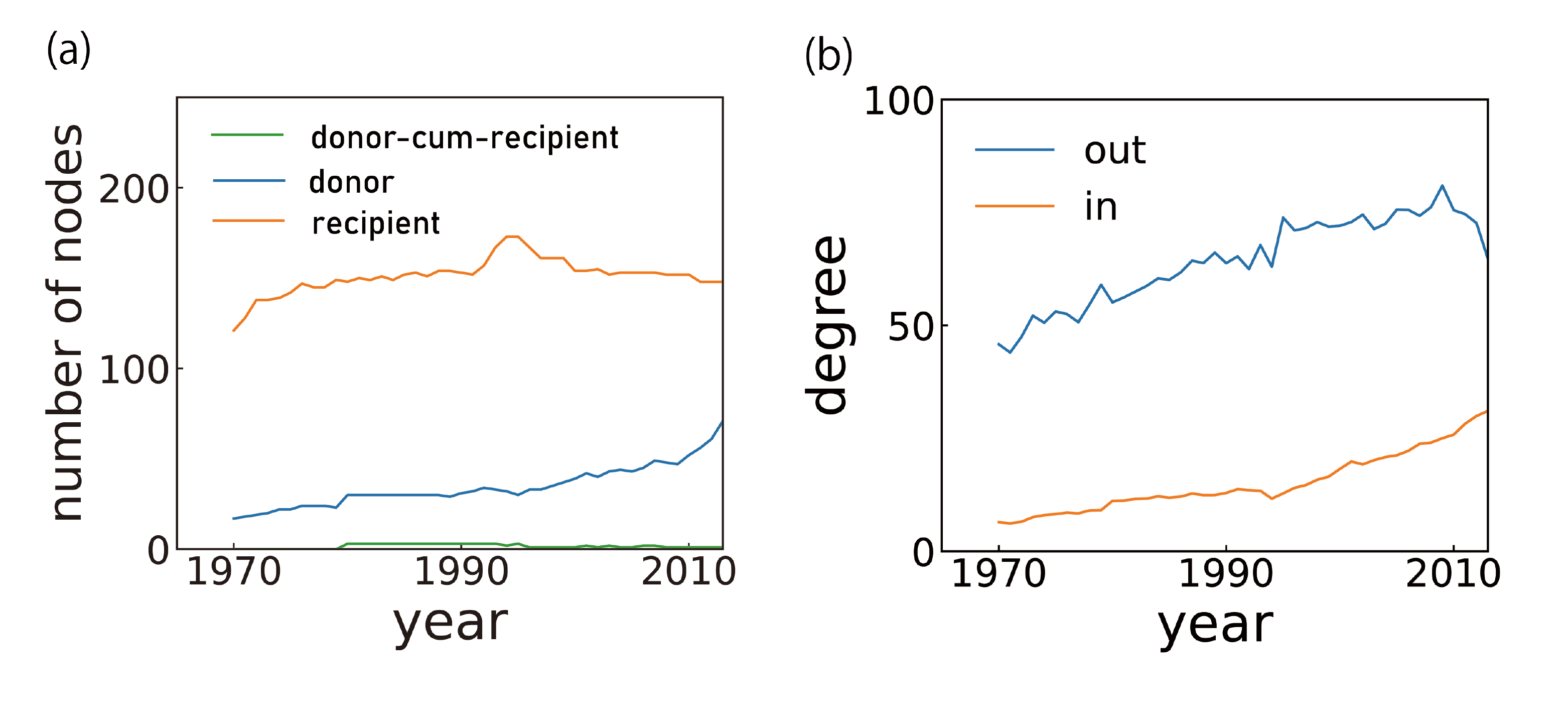}
  \caption{
    {\bf Time evolution of the local network quantities of FANs.}
    (a) The numbers of donors, recipients, and nodes with both incoming and outgoing edges (donor-cum-recipients).
    (b) The average out-degree of donors, and the average in-degree of recipients.
  }
  \label{fig:basic_stats}
\end{figure}

Before moving to more global characteristics (i.e., block structure), we first examine local network characteristics, such as the number of nodes and average degree.
Figure~\ref{fig:basic_stats} shows the growing trend of the FANs during the observation period.
As shown in Fig~\ref{fig:basic_stats}(a), the numbers of donors and recipients steadily increase from the 1970s to the 1990s,
where donors and recipients are defined as the nodes with outgoing and incoming edges, respectively.

The number of recipients temporarily increased after the end of the Cold War and then quickly declined. The increase was due to the birth of new states created by the dissolution of the Soviet Union and Yugoslavia with the end of the Cold War, and the subsequent decline came with the conversion of countries such as Korea and the oil-producing countries in the Middle East from recipients to donors. Thereafter, the number of recipients more or less stabilizes, while the number of donors continues to increase.
Note that most nodes are either donors or recipients. Thus, the FANs are close to bipartite networks.
However, there is a small fraction of nodes with both incoming and outgoing edges, which are represented as donor-cum-recipients in Fig~\ref{fig:basic_stats}(a).
Figure~\ref{fig:basic_stats}(b) shows the average out-degree of the donors and the average in-degree of the recipients, which again shows the increasing trend over the entire observation period.
Moreover, the average out-degree of the donors is always much larger than the average in-degree of the recipients, which indicates a small number of donors provide aid to a large number of recipients (see also S1 Fig. and S7 Table). \\

\subsection*{Time evolution of block structures}

\begin{figure}[!h]
  
    \includegraphics[width=13cm]{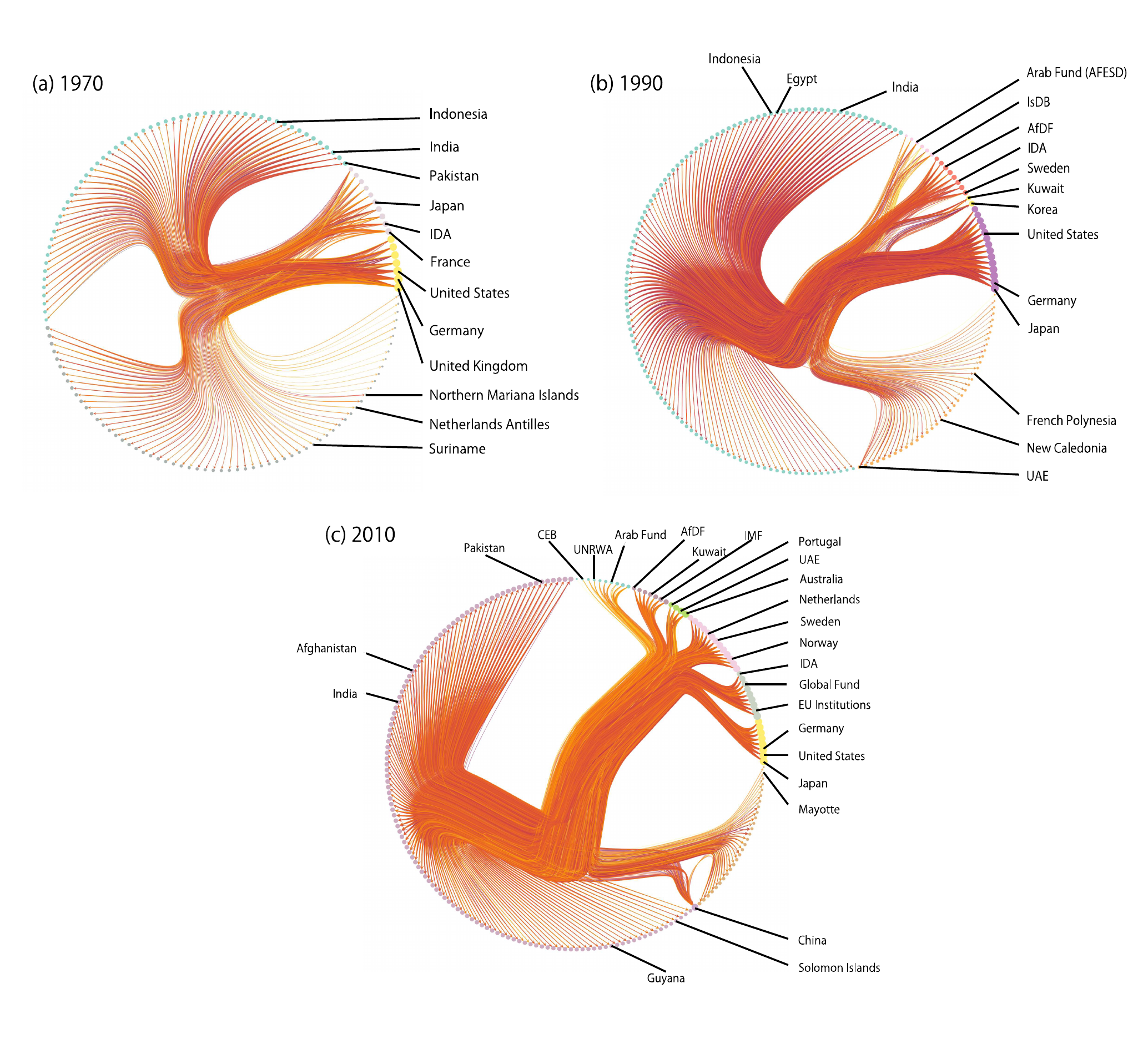}\\
  \caption{
    {\bf Block structure of FANs.}
    The block structure of FANs in (a) 1970, (b) 1990, and (c) 2010.
    Circles represent nodes whose colors indicate their blocks.
    The arrows between nodes represent edges whose colors indicate the edge weights.
    Nodes in each block are listed in Tables
    S4, S5, and S6.
  }
  \label{fig:snapshot}
\end{figure}

Next, we examine the block structures of the FANs using hierarchical SBMs in order to study larger-scale network characteristics.
Figure~\ref{fig:snapshot} shows samples of obtained block structures of foreign aid networks in 1970, 1990, and 2010.
For each FAN, we take 100 samples from the posterior probability distribution and keep only stable blocks that comprise the nodes that belong to the block in at least 95\% of the samples.
Nodes in each stable block are listed in S1, S2, and S3 Tables
.
The donors and recipients are split into different blocks, indicating that the inference of the block structure works as expected.
We also find finer structures within the donor blocks and recipient blocks.
Note that the division between donor and recipient blocks shows that the SBM inference correctly captures the fact that FANs are close to bipartite networks.
This is not the case for the inference of communities that are sets of nodes more densely connected internally as compared to externally.

\begin{figure}[!h]
 
    \includegraphics[width=13cm]{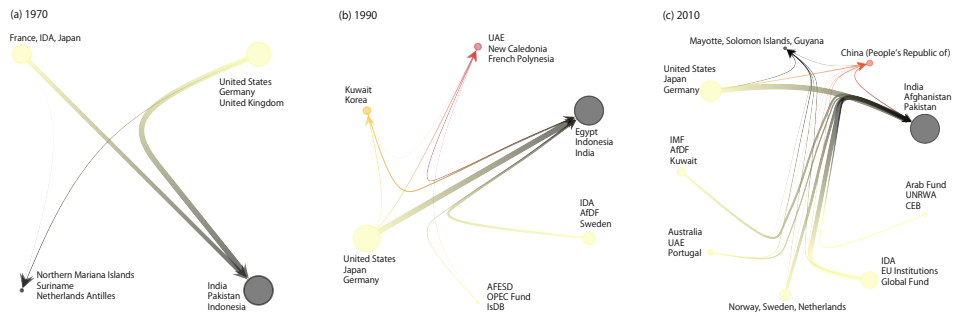}\\
  \caption{
    {\bf Aid flow networks of blocks.}
    A coarse-grained representation of the FANs in (a) 1970, (b) 1990, and (c) 2010, where the nodes in this figure represent the blocks.
    The edges show the aid flow between the nodes in each block.
    The aid flow from block $b_1$ to block $b_2$ is the total aid given by actors in $b_1$ to actors in $b_2$.
    The size of the nodes depends on the total inward and outward aid flow summed over all members of the blocks.
    The labels show the names of the actors with the three largest inward and outward aid flows in each block.
    The brightness of the blocks indicates the fractions of outward aid flow, in other words, bright blocks mainly contain donors.
    The edge width shows the volume of aid flow.
  }
  \label{fig:interblock}
\end{figure}

The division between the donor blocks and recipient blocks is also observed in Fig~\ref{fig:interblock}, which shows the networks of aid flow between the blocks.
Most of the blocks include only donors (indicated in yellow in Fig~\ref{fig:interblock}) or only recipients (black).
Meanwhile, Fig~\ref{fig:interblock} shows that the networks sometimes include a block comprising both donors and recipients (red nodes).
Note that these blocks comprise emerging donors, such as the UAE, Kuwait, Korea, and, more recently, China. \\

\begin{figure}[!h]
  
    \includegraphics[width=8cm]{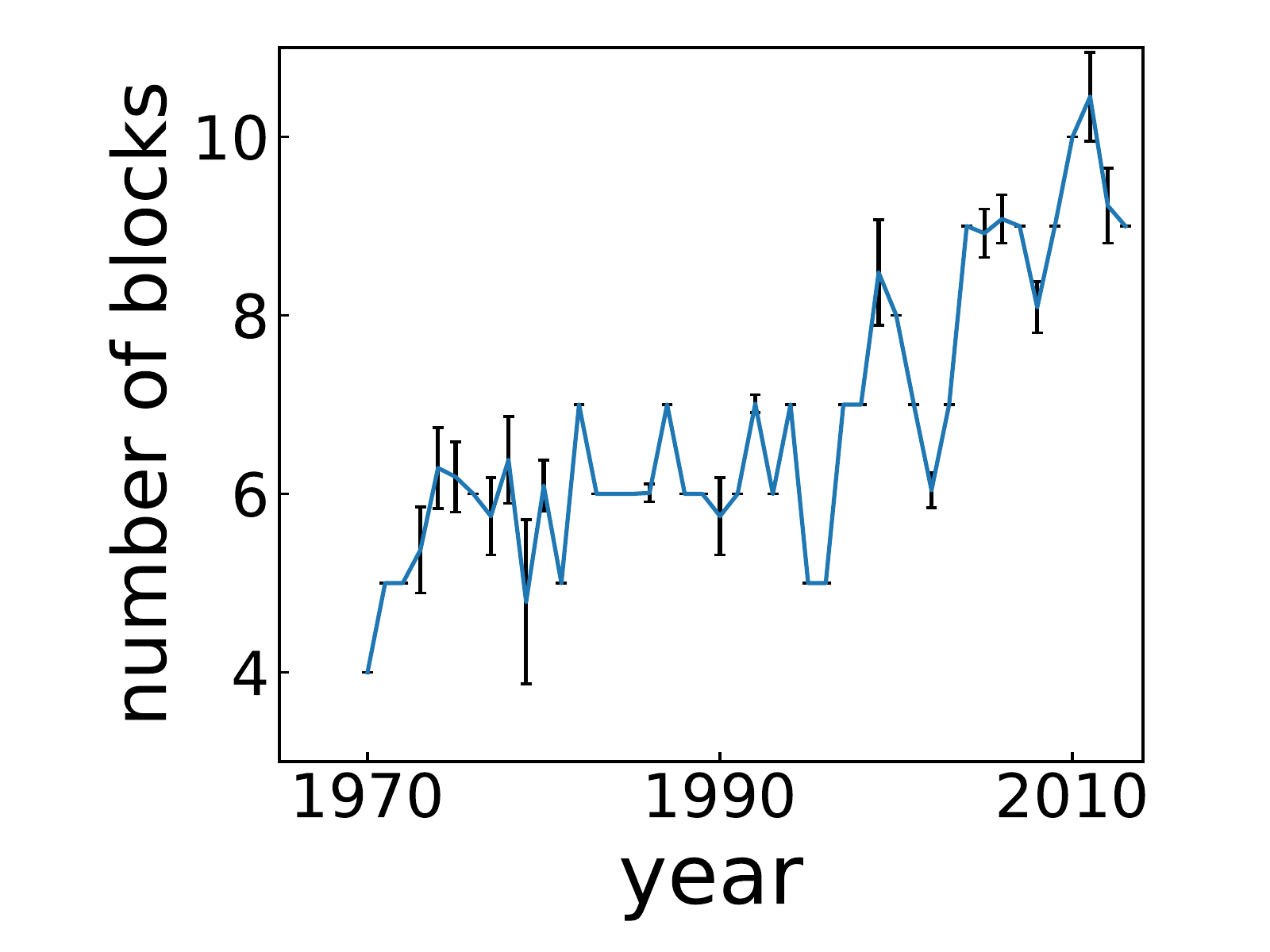}
  \caption{
    {\bf The number of blocks.}
    Averaged over 100 samples in posterior probability distribution sampling.
    Error bars show the standard deviation.
  }
  \label{fig:num_block}
\end{figure}

Both Figs~\ref{fig:snapshot} and \ref{fig:interblock} clearly show that the block structures of the FANs become more complex over time.
The FAN in 1970 consists of only four blocks while those in 1990 and 2010 have six and nine blocks, respectively.
While the recipients are mostly divided into two blocks for these networks, the number of donor blocks increases from two to seven,
indicating increasing complexity of the FANs, largely due to the divisions of the donor blocks.
We also confirm the increase in the number of blocks in Fig~\ref{fig:num_block}, which shows the number of blocks averaged over 100 samples taken from the posterior probability distribution.\\

\begin{figure}[!h]
  
    \includegraphics[width=13cm, angle = 0]{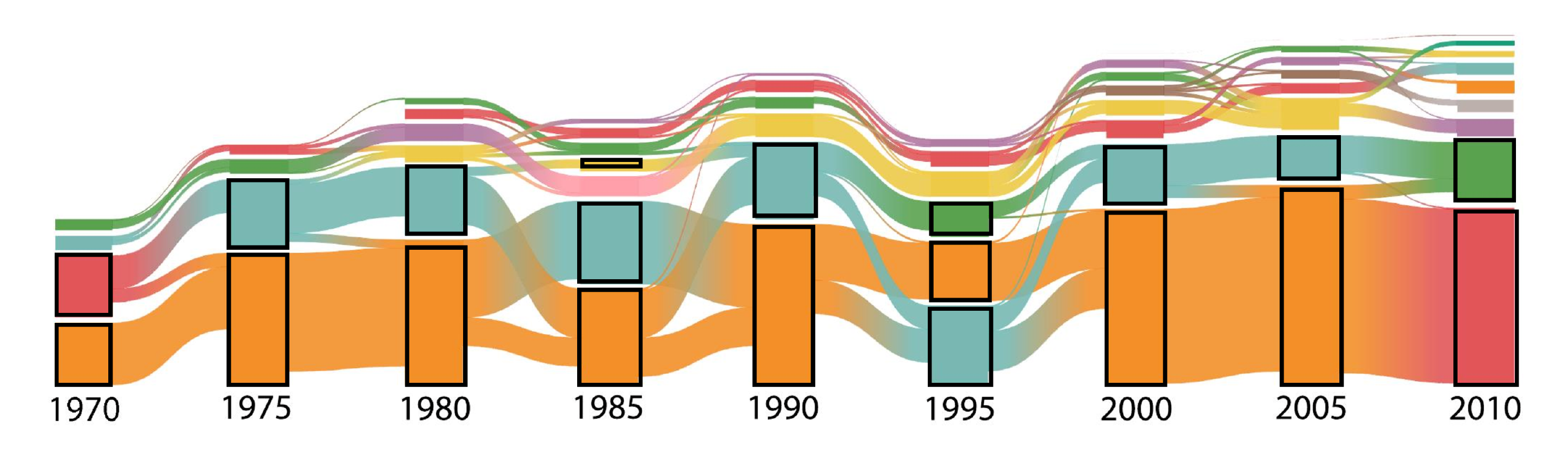}\\
    \caption{
    {\bf The alluvial diagram of stable blocks.}
    The diagram shows the change of memberships of the stable blocks from 1970 to 2010 every 5 years (from left to right).
    The flow from block $b_1$ in year $t$ to block $b_2$ in year $t+5$ represents a situation in which actors who had belonged to $b_1$ in year $t$ belonged to $b_2$ in year $t+5$.
    The height of each block indicates the number of nodes (actors) in the block.
    Colors are used to make blocks distinguishable and do not represent any block characteristics.
    The blocks surrounded by black lines are the recipient blocks.
    }
    \label{fig:alluvial}
\end{figure}

We also find that the time evolution of the donor block structures is different from those of the recipients.
Figure~\ref{fig:alluvial} shows the change in the membership of blocks between different time points (alluvial diagram generated by \cite{alluvial}).
In most of the time points, the two bottom blocks are recipients while the top blocks are donors.
Although the two or three recipient blocks occasionally change their membership (e.g., in 1985 and 1995), recipient blocks are more stable than the donor blocks (see also S2 Fig.).
Meanwhile, the donor blocks frequently split and change their members, which largely contributes to the increase in the number of blocks in the observation period.
Despite these unsettling dynamics, we note that the most resourceful block in terms of total aid commitment has almost always had a stable set of powerful donors at its core, including the U.S., Japan, and Germany (see Fig~\ref{fig:interblock} and S7 Table).\\

\subsection*{Characterization of blocks}
\begin{figure}[!h]

  \includegraphics[width=13cm]{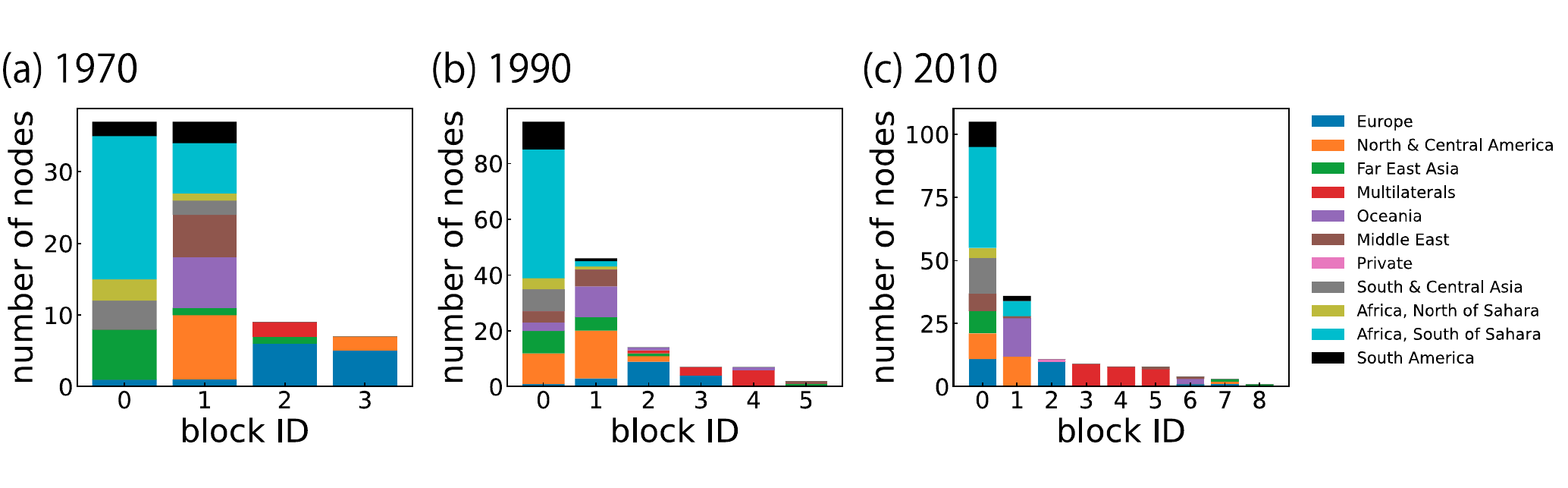}\\
  \caption{
    {\bf Regions or types in FAN stable blocks in (a) 1970, (b) 1990, and (c) 2010.}
    The number of actors of each region or type in each block.
    Governments are labeled by the regions of their countries, while IGOs and NGOs are labeled as multilateral and private, respectively.
    The labels are available from OECD data.
    The stable blocks are in descending order of size.
  }
  \label{fig:composition_region}
\end{figure}

We also analyze the characteristics of each block in the FANs.
Figure~\ref{fig:composition_region} shows the regions or types of actors in the blocks in 1970, 1990, and 2010, indicating that each block consists of similar nodes in terms of the region or type.
For example, most sub-Saharan African countries and Oceanic countries belong to the same block (i.e., Blocks 1 and 2, respectively).
This means that the nodes (especially recipients) in the same region tend to have similar connection patterns.
Moreover, multilateral donors tend to form their own blocks: Block 4 in 1990 and Blocks 3--5 in
2010.
This finding means that multilateral donors have different patterns of connections to those of the bilateral donors and there is a variety of such patterns, even among multilateral donors.\\

\begin{figure}[!h]
  
  \includegraphics[width=13cm]{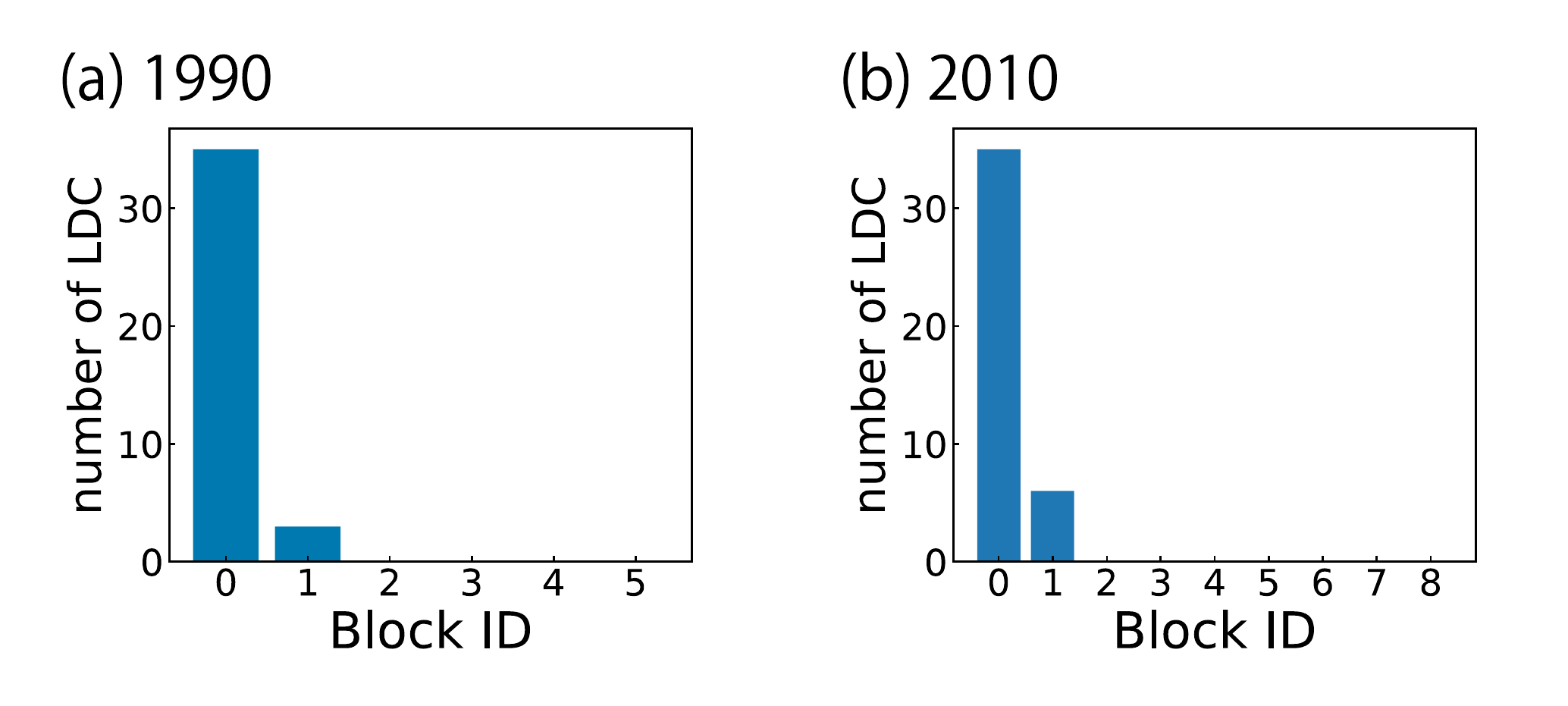}\\
  \caption{
    {\bf The number of LDCs in each stable block in (a) 1990 and (b) 2010.}
    The stable blocks are in descending order of size.
  }
  \label{fig:ldc_fraction}
\end{figure}

Next, we examine the relationship between the poverty of the nodes and the block structure of the FANs.
Figure~\ref{fig:ldc_fraction} shows the number of least developed countries (LDCs) in each block in 1990 and 2010 (LDC data are not available for 1970) \cite{LDClist}.
As shown in the figure, most LDCs are included in the largest block (Block 0), indicating that LDCs have received a similar pattern of the development aid.\\

\section*{Discussion and conclusions}
We constructed FANs for 1970--2013 and studied their local and global properties to examine the evolving structure of global development cooperation among numerous international actors. The results of this largely exploratory investigation corroborate several important observations and insights. First, the global structure of aid flows, having been continuously incorporating new participants, especially on the donor side, has steadily expanded and become more complex over time (see Fig~\ref{fig:num_block}). This largely confirms the prevailing view of aid relations as an ever-expanding domain of IR \cite{brown2020rise,heldt2019explaining,hook2016development}.

Second, and despite the former conclusion, the global aid-flow structure, at its core, has largely stable components (see Fig~\ref{fig:interblock}): namely, a limited number of lead donors, such as the U.S., Japan, and the EU, which are financially connected, to a varying degree, to an entire array of recipient groups (and even some donor groups); and a large number of poor recipients, including most sub-Saharan African countries, which are well connected to a broad array of donor groups, especially the lead donor group. This observation may be somewhat surprising given recurrent efforts for ``aid reform'' during the past decades, which have been aimed at bringing meaningful changes to global aid flow patterns. Of course, changes have occurred, most notably, the expanding roster of multilateral and bilateral donors with accompanying diversification in aid-flow channels. In comparison with the long-established lead donors, however, the roles of these other donors remain complementary, if not peripheral.

Third, for a large part of the study period, especially since the 1980s, we observed the recurrent formation of small but distinct donor-cum-recipient blocks comprising emerging countries, such as Kuwait, Korea, and Turkey.
Since the 2000s, unlike the previous donor-cum-recipient blocks, the block tends to comprise only China (although it sometimes comprises other emergent donors, e.g., Turkey).
Furthermore, China's block, while being deeply embedded in the global aid structure, is highly distinct and persistent.
In comparison with other leading donors, however, its presence remains limited. This is probably because the AidData dataset covers the aid given by China only for the period from 2000 to 2013, which is largely before the launch of China's massive Belt and Road Initiative.

There are numerous limitations of our study, which we will continue to work on in the immediate future. First, it is necessary to iterate the present line of analysis in different modeling specifications to establish the robustness of our findings. Data availability, as just mentioned, is another obvious challenge. Fortunately, in this regard, AidData has recently released a new version of the China dataset, which extends its temporal coverage beyond 2013 to 2017. Incorporating a more recent part of China's aid behavior into the analysis might substantially change its standing in the derived block configurations from the modest contribution that we have seen so far.

Finally, the present analysis is exploratory as well as descriptive. We believe that such an endeavor is an essential step to follow, but we also recognize the need to go beyond it.
A useful next step would be to establish some theoretical baseline for understanding and evaluating the obtained results, since currently, there are not obvious conclusions about the extent to which a given global structure of aid relations is effective in attaining coordination among numerous donors, aligning development efforts between donors and recipients, or delivering needed resources to intended beneficiaries, etc. We aim to expand our inquiry into these more challenging domains by again employing various modeling devices provided by network science (e.g., exponential random graph models).

%
%
%
\bibliographystyle{plos2015}
\bibliography{main}

\renewcommand{\thetable}{S\arabic{table}}

\clearpage
\section*{Supporting Information}

\begin{figure}[!h]
  \includegraphics[width=10cm]{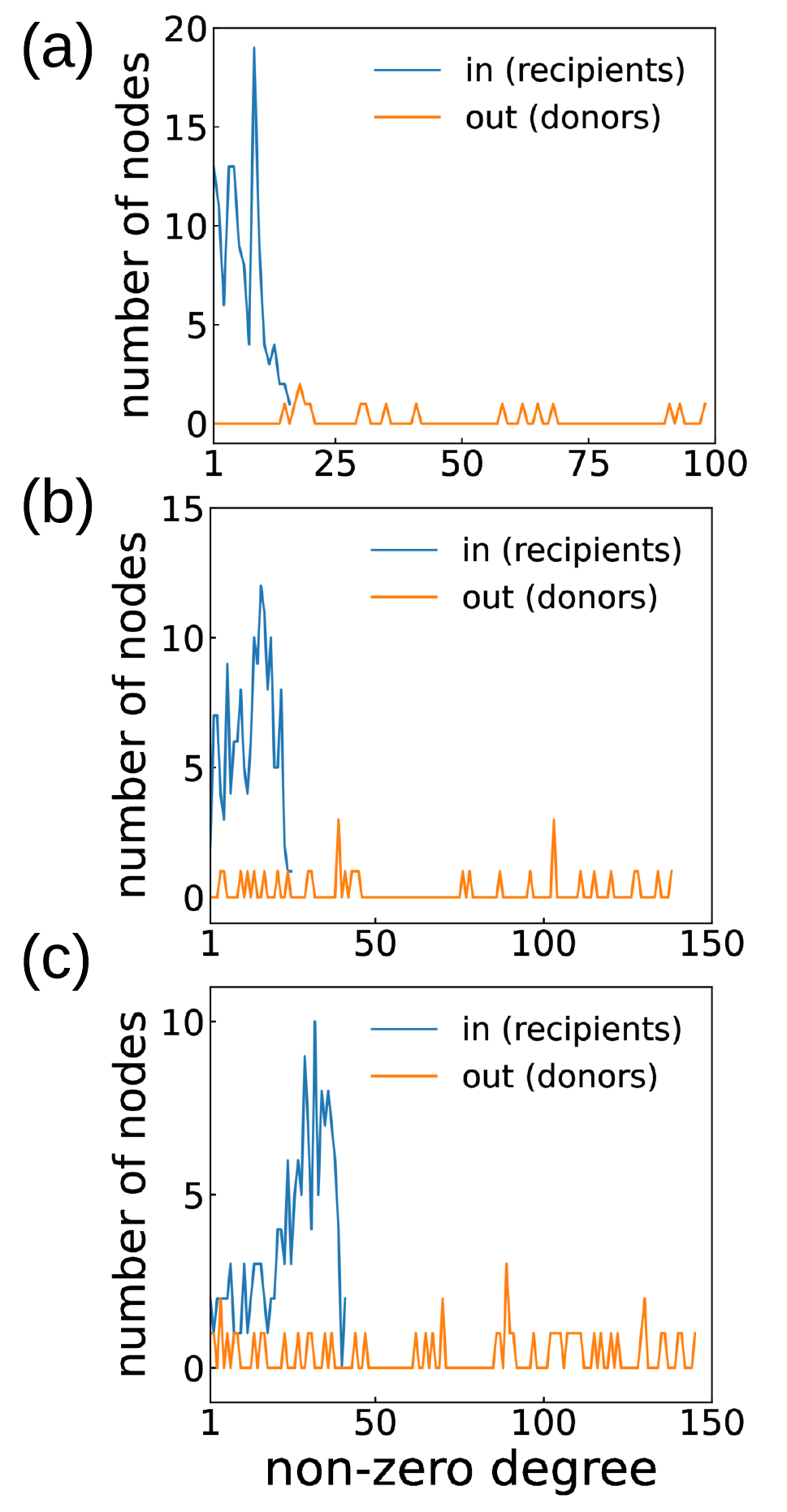}
  \renewcommand\thefigure{}
  \caption{
      {\bf S1: In-degree histogram of recipients and out-degree histogram of donors in (a) 1970, (b) 1990, and (c) 2010.}
  }
  \label{fig:degree_dist}
\end{figure} 

\begin{figure}[!h]
  \includegraphics[width=10cm]{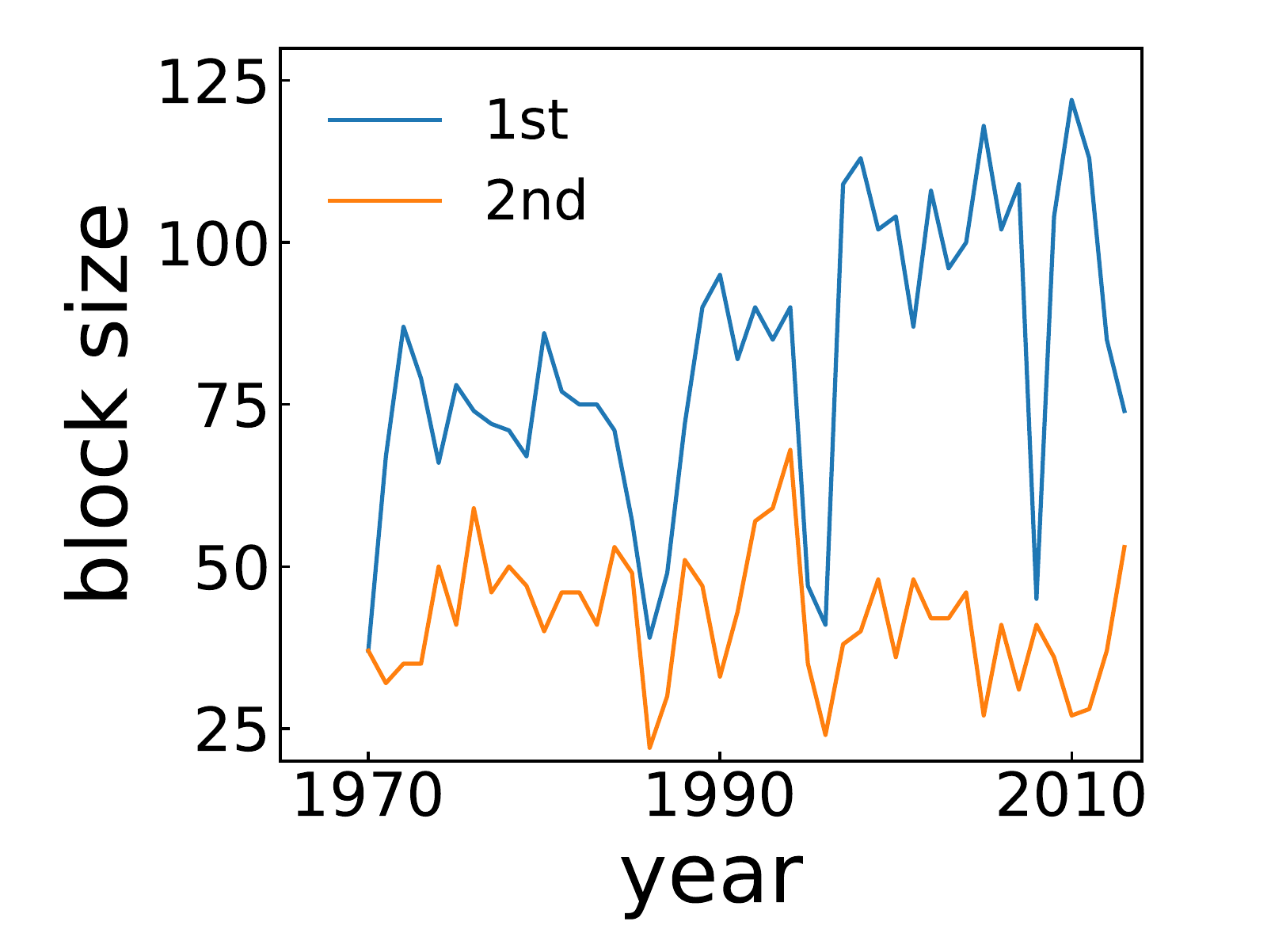}\\
  \renewcommand\thefigure{}
  \caption{
    {\bf S2: Time evolution of the sizes of the largest and second largest stable blocks.} 
    }
  \label{fig:block_size}
\end{figure}

\begin{table}[!ht]
\centering
\caption{
{\bf List of actors in stable blocks in 1970.}}
\begin{tabular}{lp{11cm}}
\hline
block ID & actors\\
\hline
0 & Afghanistan, Algeria, Botswana, Brazil, Burkina Faso, Burundi, Cambodia, Cameroon, Central African Republic, Congo, Democratic Republic of the Congo, Ethiopia, Gabon, Ghana, India, Indonesia, Kenya, Korea, Lao People's Democratic Republic, Madagascar, Mali, Morocco, Niger, Nigeria, Pakistan, Peru, Philippines, Rwanda, Senegal, Sri Lanka, Tanzania, Thailand, Tunisia, Turkey, Uganda, Viet Nam, Zambia\\
1 & Australia, Benin, Bolivia, Chad, Chile, Chinese Taipei, Colombia, Costa Rica, Cote d'Ivoire, Cyprus, Dominican Republic, Ecuador, Egypt, El Salvador, Eswatini, Fiji, Gambia, Guatemala, Guinea, Guyana, Honduras, Hong Kong (China), Iran, Israel, Jordan, Lebanon, Lesotho, Liberia, Malawi, Malaysia, Malta, Mauritania, Mauritius, Mexico, Myanmar, Nepal, New Caledonia, Nicaragua, Panama, Papua New Guinea, Sierra Leone, Singapore, Somalia, Sudan, Togo, Uruguay, Vanuatu, Venezuela\\
2 & Angola, Argentina, Bahamas, Bahrain, Barbados, Belize, Bermuda, Bhutan, Brunei Darussalam, Comoros, Cuba, Djibouti, French Polynesia, Gibraltar, Haiti, Iraq, Jamaica, Kiribati, Libya, Maldives, Mozambique, Netherlands Antilles, Northern Mariana Islands, Paraguay, Saint Helena, Samoa, Saudi Arabia, Seychelles, Solomon Islands, Suriname, Syrian Arab Republic, Tonga, Trinidad and Tobago, United Arab Emirates, Wallis and Futuna, Yemen, Zimbabwe\\
3 & Austria, Denmark, EU Institutions, France, International Development Association [IDA], Japan, Netherlands, Norway, Sweden\\
4 & Belgium, Canada, Germany, Italy, Switzerland, United Kingdom, United States\\
\hline
\end{tabular}
\label{tab:list_1970}
\end{table}

\begin{table}[!ht]
\centering
\caption{
{\bf List of actors in stable blocks in 1990.}}
\begin{tabular}{lp{11cm}}
\hline
block ID & actors\\
\hline
0 & Afghanistan, Algeria, Angola, Argentina, Bangladesh, Benin, Bhutan, Bolivia, Botswana, Brazil, Burkina Faso, Burundi, Cabo Verde, Cambodia, Cameroon, Central African Republic, Chad, Chile, China (People's Republic of), Colombia, Comoros, Congo, Costa Rica, Cote d'Ivoire, Cuba, Democratic Republic of the Congo, Djibouti, Dominican Republic, Ecuador, Egypt, El Salvador, Equatorial Guinea, Eswatini, Ethiopia, Gabon, Gambia, Ghana, Guatemala, Guinea, Guinea-Bissau, Guyana, Haiti, Honduras, India, Indonesia, Iran, Jamaica, Jordan, Kenya, Lao People's Democratic Republic, Lebanon, Lesotho, Liberia, Madagascar, Malawi, Malaysia, Maldives, Mali, Mauritania, Mauritius, Mexico, Morocco, Mozambique, Namibia, Nepal, Nicaragua, Niger, Nigeria, Pakistan, Papua New Guinea, Paraguay, Peru, Philippines, Rwanda, Saint Lucia, Samoa, Sao Tome and Principe, Senegal, Seychelles, Sierra Leone, Somalia, Sri Lanka, Sudan, Tanzania, Thailand, Togo, Tonga, Tunisia, Turkey, Uganda, Uruguay, Viet Nam, Yemen, Zambia, Zimbabwe\\
1 & Albania, Anguilla, Antigua and Barbuda, Aruba, Bahamas, Bahrain, Barbados, Belize, Bermuda, British Virgin Islands, Brunei Darussalam, Cayman Islands, Chinese Taipei, Cook Islands, Cyprus, Democratic People's Republic of Korea, Dominica, French Polynesia, Gibraltar, Grenada, Iraq, Kiribati, Libya, Macau (China), Mayotte, Mongolia, Montserrat, Nauru, Netherlands Antilles, New Caledonia, Niue, Oman, Qatar, Saint Helena, Saint Kitts and Nevis, Saint Vincent and the Grenadines, Saudi Arabia, Solomon Islands, Suriname, Tokelau, Trinidad and Tobago, Turks and Caicos Islands, Tuvalu, United Arab Emirates, Vanuatu, Wallis and Futuna\\
2 & Australia, Austria, Belgium, Canada, EU Institutions, Finland, France, Germany, Italy, Japan, Netherlands, Switzerland, United Kingdom, United States\\
3 & Fiji, Hong Kong (China), Israel, Malta, Myanmar, Northern Mariana Islands, Panama, Singapore, Syrian Arab Republic, Venezuela\\
4 & African Development Fund [AfDF], Denmark, IFAD, International Development Association [IDA], Ireland, Norway, Sweden\\
5 & Arab Bank for Economic Development in Africa [BADEA], Arab Fund (AFESD), Caribbean Development Bank [CarDB], Islamic Development Bank [IsDB], New Zealand, Nordic Development Fund [NDF], OPEC Fund for International Development [OPEC Fund]\\
6 & Korea, Kuwait\\
\hline
\end{tabular}
\label{tab:list_1990}
\end{table}

\begin{table}[!ht]
\centering
\caption{
  {\bf List of actors in stable blocks in 2010.}}
\begin{tabular}{lp{11cm}}
\hline
block ID & actors\\
\hline
0 & Afghanistan, Albania, Algeria, Angola, Argentina, Armenia, Azerbaijan, Bangladesh, Belarus, Benin, Bolivia, Bosnia and Herzegovina, Botswana, Brazil, Burkina Faso, Burundi, Cambodia, Cameroon, Central African Republic, Chad, Chile, Colombia, Congo, Costa Rica, Cote d'Ivoire, Croatia, Cuba, Democratic Republic of the Congo, Djibouti, Dominican Republic, Ecuador, Egypt, El Salvador, Eswatini, Ethiopia, Gambia, Georgia, Ghana, Guatemala, Guinea, Guinea-Bissau, Haiti, Honduras, India, Indonesia, Iran, Iraq, Jordan, Kazakhstan, Kenya, Kosovo, Kyrgyzstan, Lao People's Democratic Republic, Lebanon, Lesotho, Liberia, Madagascar, Malawi, Malaysia, Mali, Mauritania, Mexico, Moldova, Mongolia, Montenegro, Morocco, Mozambique, Myanmar, Namibia, Nepal, Nicaragua, Niger, Nigeria, North Macedonia, Pakistan, Palestinian Adm. Areas, Panama, Paraguay, Peru, Philippines, Rwanda, Senegal, Serbia, Sierra Leone, Somalia, South Africa, Sri Lanka, Sudan, Syrian Arab Republic, Tajikistan, Tanzania, Thailand, Togo, Tunisia, Turkey, Turkmenistan, Uganda, Ukraine, Uruguay, Uzbekistan, Venezuela, Viet Nam, Yemen, Zambia, Zimbabwe\\
1 & Anguilla, Antigua and Barbuda, Barbados, Belize, Cook Islands, Dominica, Equatorial Guinea, Fiji, Grenada, Guyana, Jamaica, Kiribati, Marshall Islands, Mauritius, Mayotte, Micronesia, Montserrat, Nauru, Niue, Oman, Palau, Saint Helena, Saint Kitts and Nevis, Saint Lucia, Saint Vincent and the Grenadines, Samoa, Sao Tome and Principe, Seychelles, Solomon Islands, Suriname, Tokelau, Tonga, Trinidad and Tobago, Tuvalu, Vanuatu, Wallis and Futuna\\
2 & Bhutan, Cabo Verde, Canada, Comoros, Democratic People's Republic of Korea, Eritrea, Finland, France, Gabon, Ireland, Korea, Libya, Maldives, Papua New Guinea, Spain, Timor-Leste, UN Peacebuilding Fund [UNPBF], United Kingdom\\
3 & Austria, Belgium, Bill \& Melinda Gates Foundation, Denmark, Greece, Italy, Luxembourg, Netherlands, Norway, Sweden, Switzerland\\
4 & EU Institutions, Global Alliance for Vaccines and Immunization [GAVI], Global Environment Facility [GEF], Global Fund, International Development Association [IDA], UNAIDS, UNDP, UNFPA, UNICEF\\
5 & Adaptation Fund, African Development Bank [AfDB], Arab Fund (AFESD), Council of Europe Development Bank [CEB], Nordic Development Fund [NDF], OSCE, UNECE, UNRWA\\
6 & African Development Fund [AfDF], Arab Bank for Economic Development in Africa [BADEA], Caribbean Development Bank [CarDB], IFAD, IMF (Concessional Trust Funds), Islamic Development Bank [IsDB], Kuwait, OPEC Fund for International Development [OPEC Fund]\\
7 & Australia, New Zealand, Portugal, United Arab Emirates\\
8 & Germany, Japan, United States\\
9 & China (People's Republic of)\\
\hline
\end{tabular}
\label{tab:list_2010}
\end{table}

\begin{table}[!ht]
\centering
\caption{
  {\bf List of actors in the sample block structure in 1970 in Fig. \ref{fig:snapshot}}
}
\begin{tabular}{lp{11cm}}
\hline
block ID & actors\\
\hline
0 & Angola, Argentina, Bahamas, Bahrain, Barbados, Belize, Benin, Bermuda, Bhutan, Brunei Darussalam, Chile, Chinese Taipei, Colombia, Comoros, Costa Rica, Cuba, Cyprus, Djibouti, Dominican Republic, El Salvador, Fiji, French Polynesia, Gibraltar, Guinea, Guyana, Haiti, Honduras, Hong Kong (China), Iraq, Israel, Jamaica, Jordan, Kiribati, Lebanon, Liberia, Libya, Maldives, Malta, Mauritania, Mozambique, Netherlands Antilles, New Caledonia, Nicaragua, Northern Mariana Islands, Panama, Paraguay, Saint Helena, Samoa, Saudi Arabia, Seychelles, Singapore, Solomon Islands, Somalia, Sudan, Suriname, Syrian Arab Republic, Tonga, Trinidad and Tobago, United Arab Emirates, Uruguay, Vanuatu, Venezuela, Wallis and Futuna, Yemen, Zimbabwe\\
1 & Afghanistan, Algeria, Bolivia, Botswana, Brazil, Burkina Faso, Burundi, Cambodia, Cameroon, Central African Republic, Chad, Congo, Cote d'Ivoire, Democratic Republic of the Congo, Ecuador, Egypt, Eswatini, Ethiopia, Gabon, Gambia, Ghana, Guatemala, India, Indonesia, Iran, Kenya, Korea, Lao People's Democratic Republic, Lesotho, Madagascar, Malawi, Malaysia, Mali, Mauritius, Mexico, Morocco, Myanmar, Nepal, Niger, Nigeria, Pakistan, Papua New Guinea, Peru, Philippines, Rwanda, Senegal, Sierra Leone, Sri Lanka, Tanzania, Thailand, Togo, Tunisia, Turkey, Uganda, Viet Nam, Zambia\\
2 & Australia, Austria, Denmark, EU Institutions, France, International Development Association [IDA], Japan, Netherlands, Norway, Sweden\\
3 & Belgium, Canada, Germany, Italy, Switzerland, United Kingdom, United States\\
\hline
\end{tabular}
\label{tab:sample_1970}
\end{table}

\begin{table}[!ht]
\centering
\caption{
  {\bf List of actors in the sample block structure in 1990 in Fig. \ref{fig:snapshot}}
}
\begin{tabular}{lp{11cm}}
\hline
block ID & actors\\
\hline
0 & Afghanistan, Algeria, Angola, Argentina, Bahrain, Bangladesh, Benin, Bhutan, Bolivia, Botswana, Brazil, Brunei Darussalam, Burkina Faso, Burundi, Cabo Verde, Cambodia, Cameroon, Central African Republic, Chad, Chile, China (People's Republic of), Chinese Taipei, Colombia, Comoros, Congo, Costa Rica, Cote d'Ivoire, Cuba, Democratic Republic of the Congo, Djibouti, Dominican Republic, Ecuador, Egypt, El Salvador, Equatorial Guinea, Eswatini, Ethiopia, Fiji, Gabon, Gambia, Ghana, Grenada, Guatemala, Guinea, Guinea-Bissau, Guyana, Haiti, Honduras, Hong Kong (China), India, Indonesia, Iran, Israel, Jamaica, Jordan, Kenya, Lao People's Democratic Republic, Lebanon, Lesotho, Liberia, Libya, Madagascar, Malawi, Malaysia, Maldives, Mali, Malta, Mauritania, Mauritius, Mexico, Mongolia, Morocco, Mozambique, Myanmar, Namibia, Nepal, Nicaragua, Niger, Nigeria, Northern Mariana Islands, Oman, Pakistan, Panama, Papua New Guinea, Paraguay, Peru, Philippines, Rwanda, Saint Lucia, Samoa, Sao Tome and Principe, Saudi Arabia, Senegal, Seychelles, Sierra Leone, Singapore, Somalia, Sri Lanka, Sudan, Syrian Arab Republic, Tanzania, Thailand, Togo, Tonga, Tunisia, Turkey, Uganda, Uruguay, Venezuela, Viet Nam, Yemen, Zambia, Zimbabwe\\
1 & Albania, Anguilla, Antigua and Barbuda, Aruba, Bahamas, Barbados, Belize, Bermuda, British Virgin Islands, Cayman Islands, Cook Islands, Cyprus, Democratic People's Republic of Korea, Dominica, French Polynesia, Gibraltar, Iraq, Kiribati, Macau (China), Mayotte, Montserrat, Nauru, Netherlands Antilles, New Caledonia, Niue, Qatar, Saint Helena, Saint Kitts and Nevis, Saint Vincent and the Grenadines, Solomon Islands, Suriname, Tokelau, Trinidad and Tobago, Turks and Caicos Islands, Tuvalu, United Arab Emirates, Vanuatu, Wallis and Futuna\\
2 & Australia, Austria, Belgium, Canada, EU Institutions, Finland, France, Germany, Italy, Japan, Netherlands, Switzerland, United Kingdom, United States\\
3 & African Development Fund [AfDF], Denmark, IFAD, International Development Association [IDA], Ireland, Norway, OPEC Fund for International Development [OPEC Fund], Sweden\\
4 & Arab Bank for Economic Development in Africa [BADEA], Arab Fund (AFESD), Caribbean Development Bank [CarDB], Islamic Development Bank [IsDB], New Zealand, Nordic Development Fund [NDF]\\
5 & Korea, Kuwait\\
\hline
\end{tabular}
\label{tab:sample_1990}
\end{table}

\begin{table}[!ht]
\centering
\caption{
  {\bf List of actors in the sample block structure in 2010 in Fig. \ref{fig:snapshot}}
}
\begin{tabular}{lp{11cm}}
\hline
block ID & actors\\
\hline
0 & Afghanistan, Albania, Algeria, Angola, Argentina, Armenia, Azerbaijan, Bangladesh, Belarus, Belize, Benin, Bhutan, Bolivia, Bosnia and Herzegovina, Botswana, Brazil, Burkina Faso, Burundi, Cabo Verde, Cambodia, Cameroon, Central African Republic, Chad, Chile, Colombia, Comoros, Congo, Costa Rica, Cote d'Ivoire, Croatia, Cuba, Democratic People's Republic of Korea, Democratic Republic of the Congo, Djibouti, Dominican Republic, Ecuador, Egypt, El Salvador, Eritrea, Eswatini, Ethiopia, Gabon, Gambia, Georgia, Ghana, Grenada, Guatemala, Guinea, Guinea-Bissau, Guyana, Haiti, Honduras, India, Indonesia, Iran, Iraq, Jamaica, Jordan, Kazakhstan, Kenya, Kosovo, Kyrgyzstan, Lao People's Democratic Republic, Lebanon, Lesotho, Liberia, Libya, Madagascar, Malawi, Malaysia, Maldives, Mali, Mauritania, Mauritius, Mexico, Moldova, Mongolia, Montenegro, Morocco, Mozambique, Myanmar, Namibia, Nepal, Nicaragua, Niger, Nigeria, North Macedonia, Pakistan, Palestinian Adm. Areas, Panama, Papua New Guinea, Paraguay, Peru, Philippines, Rwanda, Sao Tome and Principe, Senegal, Serbia, Seychelles, Sierra Leone, Solomon Islands, Somalia, South Africa, Sri Lanka, Sudan, Syrian Arab Republic, Tajikistan, Tanzania, Thailand, Timor-Leste, Togo, Tunisia, Turkey, Turkmenistan, Uganda, Ukraine, Uruguay, Uzbekistan, Venezuela, Viet Nam, Yemen, Zambia, Zimbabwe\\
1 & Anguilla, Antigua and Barbuda, Barbados, Cook Islands, Dominica, Equatorial Guinea, Fiji, Kiribati, Marshall Islands, Mayotte, Micronesia, Montserrat, Nauru, Niue, Oman, Palau, Saint Helena, Saint Kitts and Nevis, Saint Lucia, Saint Vincent and the Grenadines, Samoa, Suriname, Tokelau, Tonga, Trinidad and Tobago, Tuvalu, Vanuatu, Wallis and Futuna\\
2 & Austria, Belgium, Bill \& Melinda Gates Foundation, Canada, Denmark, Greece, Ireland, Italy, Luxembourg, Netherlands, Norway, Sweden, Switzerland\\
3 & Adaptation Fund, African Development Bank [AfDB], Arab Fund (AFESD), Caribbean Development Bank [CarDB], Council of Europe Development Bank [CEB], Nordic Development Fund [NDF], OSCE, UN Peacebuilding Fund [UNPBF], UNECE, UNRWA\\
4 & EU Institutions, Global Alliance for Vaccines and Immunization [GAVI], Global Environment Facility [GEF], Global Fund, International Development Association [IDA], UNAIDS, UNDP, UNFPA, UNICEF\\
5 & Finland, France, Germany, Japan, Korea, Spain, United Kingdom, United States\\
6 & African Development Fund [AfDF], Arab Bank for Economic Development in Africa [BADEA], IFAD, IMF (Concessional Trust Funds), Islamic Development Bank [IsDB], Kuwait, OPEC Fund for International Development [OPEC Fund]\\
7 & Australia, New Zealand, Portugal, United Arab Emirates\\
8 & China (People's Republic of)\\
\hline
\end{tabular}
\label{tab:sample_2010}
\end{table}

\begin{table}[hbtp]  
\centering
  \renewcommand\thetable{}
  \caption{S7. Top 10 nodes for betweenness centrality}
  \label{table:betweenness_centrality}
  \begin{tabular}{llll}
    \hline
    Rank & 1970  & 1990  &  2010  \\
    \hline \hline
    1 & United Kingdom  & Japan  & Japan \\
    2 & Germany  & United Kingdom   & France \\
    3 & United States  & France  & Germany \\
    4 & Italy  &  EU Institutions  &  United States \\
    5 & Belgium & Canada & United Kingdom \\
    6 & Canada & Netherlands & UNDP \\
    7 & Switzerland & Belgium & China (People's Republic of) \\
    8 & France & United States & Korea \\
    9 & EU Institutions & Austria & EU Institutions \\
    10 & Australia & Australia & Australia \\
    \hline
  \end{tabular}
\end{table}


\end{document}